%% file: reference.tex
\documentclass[10pt,conference]{IEEEtran}
\IEEEoverridecommandlockouts
\usepackage{cite}
\usepackage{amsmath,amssymb,amsfonts}
\usepackage{algorithmic}
\usepackage{graphicx}
\usepackage{textcomp}
\usepackage{xcolor}
\usepackage{colortbl}
\usepackage{makecell}
\usepackage{subfigure}
\usepackage{color}
\usepackage{soul}

\newcommand\app{MicroHECL}

\def\BibTeX{{\rm B\kern-.05em{\sc i\kern-.025em b}\kern-.08em
    T\kern-.1667em\lower.7ex\hbox{E}\kern-.125emX}}
\begin{document}

\title{MicroHECL: High-Efficient Root Cause Localization in Large-Scale Microservice Systems}

\author{
\IEEEauthorblockN{\hspace{0.14cm}Dewei Liu\hspace{1cm}}
\IEEEauthorblockA{
\textit{\hspace{0.14cm}Fudan University\hspace{1cm}}\\
\hspace{0.14cm}Shanghai, China\hspace{1cm}}
\and
\IEEEauthorblockN{Chuan He\hspace{1cm}}
\IEEEauthorblockA{
\textit{Fudan University\hspace{1cm}}\\
Shanghai, China\hspace{1cm}}
\and
\IEEEauthorblockN{Xin Peng\hspace{1cm}}
\IEEEauthorblockA{
\textit{Fudan University\hspace{1cm}}\\
Shanghai, China\hspace{1cm}}
\and
\IEEEauthorblockN{Fan Lin\hspace{1cm}}
\IEEEauthorblockA{
\textit{Alibaba Group\hspace{1cm}}\\
Hangzhou, China\hspace{1cm}}
\and
\IEEEauthorblockN{Chenxi Zhang\hspace{1cm}}
\IEEEauthorblockA{
\textit{Fudan University\hspace{1cm}}\\
Shanghai, China\hspace{1cm}}
\and
\IEEEauthorblockN{\hspace{1cm}Shengfang Gong\hspace{1cm}}
\IEEEauthorblockA{
\textit{\hspace{1cm}Alibaba Group\hspace{1cm}}\\
\hspace{1cm}HangZhou, China\hspace{1cm}}
\and
\IEEEauthorblockN{Ziang Li\hspace{1cm}}
\IEEEauthorblockA{
\textit{Alibaba Group\hspace{1cm}}\\
HangZhou, China\hspace{1cm}}
\and
\IEEEauthorblockN{Jiayu Ou\hspace{1cm}}
\IEEEauthorblockA{
\textit{Alibaba Group\hspace{1cm}}\\
HangZhou, China\hspace{1cm}}
\and
\IEEEauthorblockN{Zheshun Wu\hspace{1cm}}
\IEEEauthorblockA{
\textit{Alibaba Group\hspace{1cm}}\\
HangZhou, China\hspace{1cm}}
}

\maketitle

\begin{abstract}
Availability issues of industrial microservice systems (e.g., drop of successfully placed orders and processed transactions) directly affect the running of the business.
These issues are usually caused by various types of service anomalies which propagate along service dependencies.
Accurate and high-efficient root cause localization is thus a critical challenge for large-scale industrial microservice systems.
Existing approaches use service dependency graph based analysis techniques to automatically locate root causes.
However, these approaches are limited due to their inaccurate detection of service anomalies and inefficient traversing of service dependency graph.
In this paper, we propose a high-efficient root cause localization approach for availability issues of microservice systems, called \app.
Based on a dynamically constructed service call graph, \app~analyzes possible anomaly propagation chains, and ranks candidate root causes based on correlation analysis.
We combine machine learning and statistical methods and design customized models for the detection of different types of service anomalies (i.e., performance, reliability, traffic).
To improve the efficiency, we adopt a pruning strategy to eliminate irrelevant service calls in anomaly propagation chain analysis.
Experimental studies show that \app~significantly outperforms two state-of-the-art baseline approaches in terms of both accuracy and efficiency.
\app~has been used in Alibaba and achieves a top-3 hit ratio of 68\% with root cause localization time reduced from 30 minutes to 5 minutes.

\end{abstract}


\section{Introduction}
\input{introduction}

\section{Background}
\input{background}

\section{\app~Overview}\label{sec:overview}
\input{overview}

\section{Anomaly Propagation Chain Analysis}\label{sec:propagation}
\input{propagation}

\section{Experimental Study}
\input{experiment}

\section{Practical Application}
\input{application}

\section{Related Work}
\input{related}

\section{Conclusion}
\input{conclusion}

\bibliographystyle{IEEEtran}
\bibliography{reference}
\end{document}

%% file: introduction.tex
Microservice architecture has been the latest trend in building cloud-native applications and more and more companies have chosen to migrate from the so-called monolithic architecture to microservice architecture~\cite{ICSA17MSTrend, Zhou_TSE18MS, Guo_FSE20Industry}.
Industrial microservice systems can include hundreds to thousands of services.
For example, the e-commerce system of Alibaba contains more than 30,000 services and includes over 300 subsystems.
Microservice systems support the core business of the companies, especially for Internet companies which provide online services through the Internet.

Availability issues have been a critical challenge for large-scale industrial microservice systems.
These systems are highly dynamic and complex.
A service can have several to thousands of instances running on different containers and dynamically created or
destroyed according to the scaling requirements at runtime;
the execution of a microservice system may involve a huge number of microservice interactions and most of them are asynchronous and involve complex invocation chains~\cite{Zhou_TSE18MS}.
In these systems, any anomaly with the quality (e.g., performance, reliability) of a service may propagate along service call chains and finally cause availability issues at the business level (e.g., drop of successfully placed orders).
When an availability issue is detected, its root cause service and anomaly type need to be located in a short time (e.g., 3 minutes) to allow the developers to fix the issue quickly.

Existing approaches cannot efficiently support the root cause localization of availability issues for large-scale microservice systems.
Developers in industry often rely on visualization tools to analyze logs and traces to identify possible anomalies and anomaly propagation chains to locate root causes~\cite{Zhou_TSE18MS, Guo_FSE20Industry}.
Researchers have proposed approaches for automatic root cause localization for microservice or the broader service-based systems using trace analysis~\cite{Cuong_traceBased, FSE2019LatentError} or service dependency graph~\cite{cscs2013monitorRank, CauseInfer2014, Hanzhang_GraphBased2019, GraphBasedformicroservice, NOMS2020MicroRCA, DCCGRID2018CloudRange, ICSOC2018Microscope}.
Trace analysis based approaches require expensive collection and processing of trace data, thus cannot efficiently work for large-scale systems.
Service dependency graph based approaches construct service dependency graphs based on service calls and causal relationships (e.g., services co-located in the same machines).
These approaches locate root causes by traversing service dependency graphs and detecting possible anomalies with the quality metrics (e.g., response time) of services.
These approaches are limited due to their inaccurate detection of service anomalies and inefficient traversing of service dependency graphs, especially when the system has many services and dependencies.

In this paper, we propose a high-efficient root cause localization approach for availability issues of microservice systems, called \app.
Given an availability issue initially reported on a service (called \textit{initial anomalous service}), \app~locates the root cause services and anomaly types (e.g., performance anomaly, reliability anomaly, and traffic anomaly) that cause the issue.
It dynamically constructs a service call graph of the target microservice system, which reflects the service dependencies and related quality metrics (e.g., response time) in the latest time window.
Based on the graph, it analyzes possible anomaly propagation chains from the initial anomalous service by traversing the graph along anomalous service calls.
To detect possible anomalies with individual service calls, we design an anomaly detection model for each of the three popular anomaly types (i.e., performance, reliability, traffic).
To improve the efficiency of anomaly propagation chain analysis, \app~adopts a pruning strategy to eliminate anomalous service call edges that are irrelevant to the current anomaly propagation chain.
Finally, \app~ranks candidate root causes based on the correlations between their quality metrics and the business metrics of the initial anomalous service.

To evaluate the effectiveness and efficiency of \app , we conduct a series of experimental studies with the data and availability issues collected from the e-commerce system of Alibaba.
The results show that \app~significantly outperforms two state-of-the-art baseline approaches MonitorRank~\cite{cscs2013monitorRank} and Microscope~\cite{ICSOC2018Microscope} in terms of both the accuracy and efficiency.
The results also confirm the effectiveness of the pruning strategy, which can significantly improve the efficiency while keeping the accuracy.

\app~has been deployed in Alibaba for more than 5 months and used to handle more than 600 availability issues.
It achieves a top-3 hit ratio of 68\% and reduces the time of typical root cause localization from 30 minutes to 5 minutes.
Feedback from the developers shows that most of them trust the recommendations of \app~and treat the recommendations as the root causes by default.
The analysis of the results also suggests some improvements of the approach.

The remainder of this paper is organized as follows.
Section II introduces background knowledge about the problem scenarios.
Section III presents an overview of \app~and Section IV details the anomaly propagation chain analysis.
Section V describes our experimental study for the evaluation of \app.
Section VI introduces the practical application of \app~in Alibaba.
Section VII discusses related work and Section VIII concludes the paper.


%% file: background.tex

The e-commerce system of Alibaba has more than 846 millions monthly active users.
It adopts the microservice architecture and contains more than 30,000 services.
These microservices are deployed with containers (i.e., Docker~\cite{docker}) and virtual machines and orchestrated by the customized orchestrator based on Kubernetes~\cite{kubernetes}, with Service Mesh~\cite{istio} applied for service communications.
The system is equipped with a large-scale monitoring infrastructure called EagleEye~\cite{EagleEye}.
EagleEye includes a tracing SDK, a real-time ETL (Extract-Transform-Load) data processing system, a batch processing computing cluster, and a web-based user interface.

As the carrier of the business, the system needs to ensure high availability.
Therefore, a business monitoring platform is deployed to raise timely alarms about availability issues.
These availability issues usually indicate problems with the running of business, for example the drop of successfully placed orders and success rate of transactions.
An availability issue can be caused by different types of anomalies, each of which is indicated by a set of metrics.
An anomaly can originate from a service and propagate along service calls, and finally cause an availability issue.
In this work, we focus on the following three types of anomalies that cause most of the availability issues in Alibaba.

\begin{itemize}
\item  \textbf{Performance Anomaly}.
Performance anomaly is indicated by anomalous increase of response time (RT).
It is usually caused by problematic implementation or improper environmental configurations (e.g., CPU/memory configurations of containers and virtual machines).

\item  \textbf{Reliability Anomaly}.
Reliability anomaly is indicated by anomalous increase of error counts (EC), i.e., the numbers of service call failures.
It is usually caused by exceptions due to code defects or environmental failures (e.g., server or network outage).

\item  \textbf{Traffic Anomaly}.
Traffic anomaly is indicated by anomalous increase or decrease of queries per second (QPS).
Anomalous traffic increase may break the services, while anomalous decrease may indicate that many requests cannot reach the services.
Traffic anomaly is usually caused by improper traffic configurations (e.g., the traffic limits of Nginx~\cite{nginx}), DoS attack, or unanticipated stress test.

\end{itemize}

In anomaly detection, the 3-sigma rule is often used to identify outliers of metric values as candidates of anomalies. 
The 3-sigma rule states that in a normal distribution almost all the values remain within three standard deviations of the mean.
The range can be represented as ($\mu- 3\sigma$, $\mu + 3\sigma$), where $\mu$ is the mean and $\sigma$ is the standard deviation.
The values within the range account for 99.73\% of all the values and the others can be regarded as outliers.


%% file: overview.tex
\app~is a high-efficient root cause localization approach for availability issues.
An availability issue is usually detected from the business perspective, e.g., the decline of successful orders.
It may be caused by different types of anomalies.
Currently \app~supports three types of anomalies, i.e., performance anomaly, reliability anomaly, and traffic anomaly.
When handling a specific type of anomalies, \app~considers a corresponding service call metric and a specific direction of anomaly propagation.
For example, when handling performance anomaly anomalies, \app~considers response time of service calls and the anomaly propagation direction from downstream to upstream.

\begin{figure}[ht]
	\centering
	\includegraphics[width=0.5\textwidth]{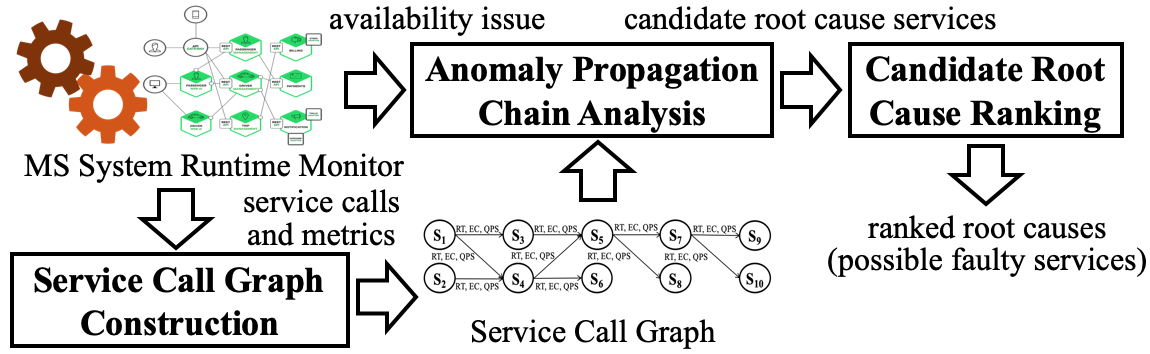}
	\vspace{-8pt}
	\caption{\app~Overview}\label{fig:overview}
	\vspace{-3pt}
\end{figure}

An overview of \app~is presented in Figure~\ref{fig:overview}.
The approach includes the following three parts.

\textbf{1) Service Call Graph Construction}

When the runtime monitor detects an availability issue, \app~starts a root cause analysis process.
It constructs a service call graph of the target microservice system based on the service calls and metrics captured by the runtime monitor.
The service call graph provides an updated snapshot of the microservice system by recording the service calls occurring in a latest time window (e.g., the last 30 minutes).
To facilitate the root cause analysis of availability issues, the service call graph also records various quality metrics of service calls, such as response time (RT), error count (EC), and queries per second (QPS).

A service call graph consists a set of service nodes S =  $\left\{ S_{1},S_{2},...,S_{n} \right\}$.
A node represents a service that is called in the latest time window (e.g., the last 30 minutes).
Note that a service may have many instances in the running system.
An edge $S_i \rightarrow S_j$ represents a service call from $S_{i}$ to $S_{j}$ that occurs in the latest time window.
For each service call, the graph records a series of values for different quality metrics (e.g., RT, EC, QPS) in the latest time window (e.g., the last 60 minutes).
For example, for each service call the graph records 60 RT values, each indicating the average response time of a minute.

The monitoring data (including service calls and their metrics) is stored in a time series database (TSDB).
When an availability issue is raised, \app~dynamically constructs a service call graph based on the data.
Services and service calls in the graph are aggregated from service instances and calls between them.
The metrics of a service call $S_i \rightarrow S_j$ are also aggregated from the corresponding metrics of the calls from the instances of $S_i$ to those of $S_j$.
It uses Flink~\cite{flink}, a distributed streaming data-flow engine, to aggregate the data.
To optimize the performance, the service call graph is constructed in an on-demand and incremental way with the anomaly propagation chain analysis process:
only when a service call is reached in the analysis the data about it is pulled from the TSDB.

\textbf{2) Anomaly Propagation Chain Analysis}

An availability issue is initially reported on a service (called \textit{initial anomalous service}), but the root cause often lies in some of its downstream or upstream services.
A root cause service and an initial anomalous service are usually connected by an anomaly propagation chain composed of a series of anomalous services.
The purpose of anomaly propagation chain analysis is to identify a set of candidate root cause services by analyzing possible anomaly propagation chains from the initial anomalous service.

Based on the service call graph, \app~analyzes possible anomaly propagation chains from the initial anomalous service of the availability issue.
The analysis is done by traversing the service call graph along anomalous service call edges, starting from the initial anomalous service and following the opposite directions of possible anomaly propagation directions.
Each anomaly propagation chain considers a specific anomaly type propagated from a neighboring service of the initial anomalous service.
To improve the efficiency, \app~adopts a pruning strategy to eliminate anomalous service call edges that are irrelevant to the current anomaly propagation chain.
This anomaly propagation chain analysis ends with a set of services as candidate root causes, each associated with a specific anomaly type.
The analysis process together with the service anomaly detection method and the pruning strategy are detailed in Section~\ref{sec:propagation}.

\textbf{3) Candidate Root Cause Ranking}

\app~ranks candidate root causes based on the possibility of causing the given availability issue.
Based on our analysis of the monitoring data, we find that the anomaly index of the initial anomalous service has similar change trends with the anomaly index of the root cause service.
Note that the anomaly index of the initial anomalous service is some kind of business metric (e.g., number of successful orders), while the anomaly index of a candidate root cause is some kind of quality metric (e.g., RT, EC, QPS).
Therefore, we use the Pearson correlation coefficient~\cite{perlationcorrelation} to measure the similarity of the change trends of the two anomaly indexes and rank the candidate root causes by the correlation coefficient.

For the initial anomalous service, the service call graph records a business metric value (e.g., number of successful orders) per minute.
For a candidate root cause service, the service call graph records a value for the quality metric (e.g., RT, EC, QPS) of the corresponding anomaly type per minute.
To reflect the latest change trends, we only consider the metric values in a latest time window (e.g., the last 60 minutes).
The values for a specific business metric of the initial anomalous service in the last $n$ minutes thus can be represented by a vector $X$, where $X_i$ ($1 \leq i \leq n$) represents the value of the $i$th minute.
Similarly, the values for a specific quality metric of a candidate root cause service in the last $n$ minutes can be represented by a vector $Y$.
Their Pearson correlation coefficient can be calculated as Equation~\ref{eq:pearson}, where $\bar{X}$ and $\bar{Y}$ represent the average of $X$ and $Y$ respectively.
A positive (negative) value of $P(X, Y)$ indicates a positive (negative) correlation; and the larger the absolute value of $P(X, Y)$ (i.e., $|P(X, Y)|$), the more likely that the two change trends correlate.
Therefore, we rank the candidate root causes by the absolute value of the Pearson correlation coefficient.

\begin{equation}
\label{eq:pearson}
P(X, Y) =  \frac{ \sum_{i=1}^{n} (X_i -\bar{X})(Y_i - \bar{Y})  )}{\sqrt{\sum_{i=1}^{n} (X_i - \bar{X})^2\sum_{i=1}^{n} (Y_i - \bar{Y})^2}}
\end{equation}


%% file: propagation.tex
In this section, we first introduce the process of anomaly propagation chain analysis, and then detail the pruning strategy and service anomaly detection methods.

\subsection{Analysis Process}\label{sec:process}
\input{process}

\subsection{Service Anomaly Detection}\label{sec:anomaly}
\input{anomaly}

\subsection{Pruning Strategy}\label{sec:pruning}
\input{pruning}

%% file: process.tex
An availability issue may be caused by different types of anomalies.
For different anomaly types the quality metrics that are considered in service anomaly detection and the anomaly propagation directions that are considered in propagation chain analysis are different.
As shown in Table~\ref{table:issue}, the considered quality metrics for performance anomaly, reliability anomaly, and traffic anomaly are response time (RT), error count (EC), and queries per second (QPS), respectively.
The propagation directions for both performance anomaly and reliability anomaly are from downstream to upstream, while the propagation direction for traffic anomaly is from upstream to downstream.

\begin{table}
	\small
	\centering
	\caption{Metrics and Propagation Direction of Availability Issues}
	\vspace{-3mm}
	\label{table:issue}
	\begin{tabular}{|c|c|c|}
		\hline
		\textbf{Anomaly Type} & \textbf{Metric} & \textbf{Propagation Direction} \\
		\hline
		Performance Anomaly & RT & downstream $\rightarrow$ upstream \\
		\hline
		Reliability Anomaly & EC & downstream $\rightarrow$ upstream \\
		\hline
		Traffic Anomaly & QPS & upstream $\rightarrow$ downstream  \\
		\hline
	\end{tabular}
	\vspace{-3mm}
\end{table}

\begin{figure}[ht]
	\centering
	\includegraphics[width=0.5\textwidth]{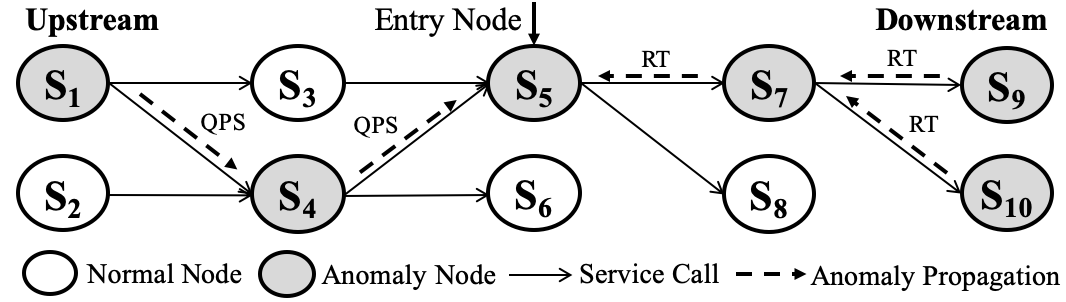}
	\vspace{-8pt}
	\caption{Anomaly Propagation Chain Analysis Process}\label{fig:propagation}
	\vspace{-5pt}
\end{figure}

The process of the anomaly propagation chain analysis is shown in Figure~\ref{fig:propagation}.
It is performed on the service call graph in the following three steps.

\textbf{1) Entry Node Analysis}.
Treat the initial anomalous service in the service call graph as the entry node.
Conduct service anomaly detection for each of its neighboring nodes (i.e., those services that directly invoke or are directly invoked by the initial anomalous service) in terms of different anomaly types and the corresponding quality metrics.
For each detected neighboring anomalous node, if its upstream/downstream relationship with the entry node is consistent with the anomaly propagation direction of the detected anomaly type, start an anomaly propagation chain analysis from it and make the detected anomaly type as the anomaly type of the anomaly propagation chain.
For example, for the entry node $S_5$ in Figure~\ref{fig:propagation}, two neighboring anomalous nodes $S_4$ and $S_7$ are detected and their anomaly types are traffic anomaly and performance anomaly respectively.
As $S_4$ is an upstream node of $S_5$ and its relationship with $S_5$ is consistent with the anomaly propagation direction of traffic anomaly (from upstream to downstream), a traffic anomaly propagation chain analysis is started from $S_4$.
Similarly, a performance anomaly propagation chain analysis is started from $S_7$.

\textbf{2) Anomaly Propagation Chain Extension}.
For each anomaly propagation chain, iteratively extend it by backtracking along the anomaly propagation direction from the starting point (i.e., a detected neighboring anomalous node of the initial anomalous service).
In each iteration, conduct service anomaly detection for all the upstream/downstream nodes of the current node in terms of the anomaly type of the current anomaly propagation chain; for each detected upstream/downstream anomalous node, add it to the current anomaly propagation chain.
The extension of the anomaly propagation chain ends when no more nodes can be added to the chain.
For example, for the anomaly propagation chain of $S_4$ the extension ends with $S_1$, which is an upstream node of $S_4$ and conforms to the propagation direction of traffic anomaly to $S_4$.
Similarity, for the anomaly propagation chain of $S_7$ the extension ends with $S_9$ and $S_{10}$.

\textbf{3) Candidate Root Causes Output}.
When all the anomaly propagation chain analysis for the initial anomalous service ends, report all the services where the extension of an anomaly propagation chain ends as candidate root causes.
For example, the candidate root causes reported for the analysis process shown in Figure~\ref{fig:propagation} include $S_1$, $S_9$, and $S_{10}$.

%% file: anomaly.tex
\begin{figure*}[ht]
    \centering
    \subfigure[RT fluctuations in two hours]{
        \includegraphics[width=5.5cm]{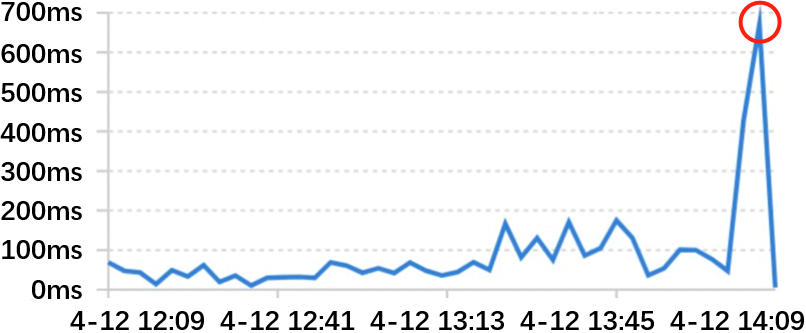}\label{fig:RT-2-hour}
    }
    \subfigure[EC fluctuations in two hours]{
        \includegraphics[width=5.5cm]{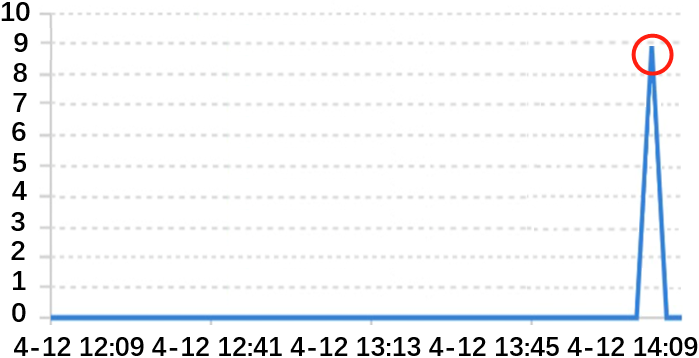}\label{fig:EC-2-hour}
    }
    \subfigure[QPS fluctuations in two hours]{
        \includegraphics[width=5.5cm]{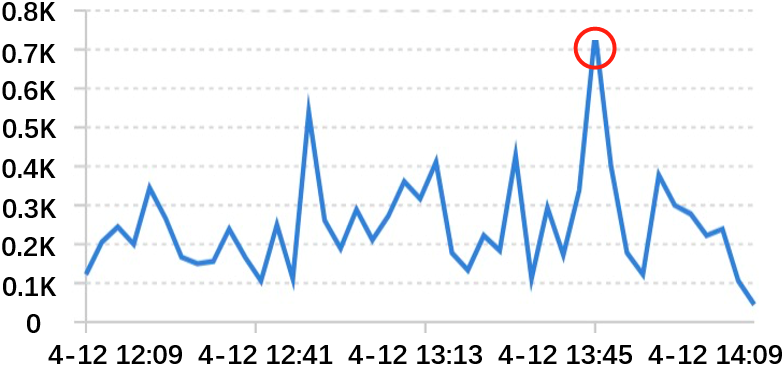}\label{fig:QPS-2-hour}
    }
    \quad
    \subfigure[RT fluctuations in one week]{
        \includegraphics[width=5.5cm]{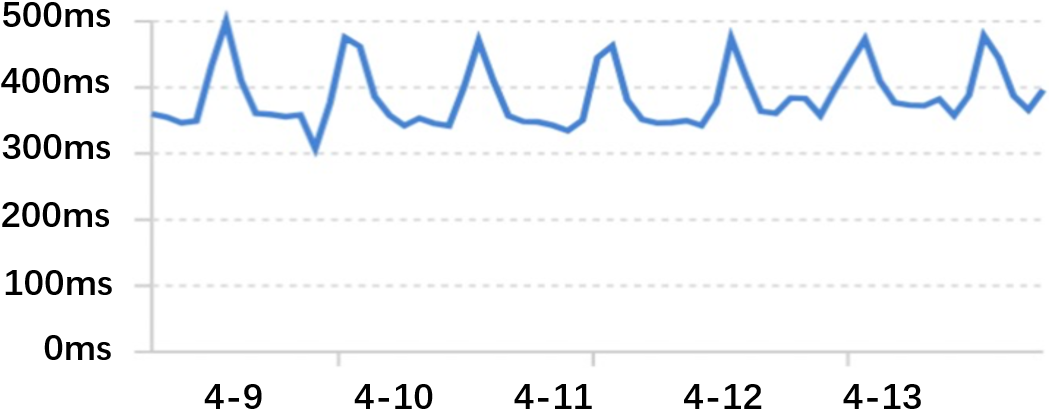}\label{fig:RT-week}
    }
    \subfigure[EC fluctuations in one week]{
        \includegraphics[width=5.5cm]{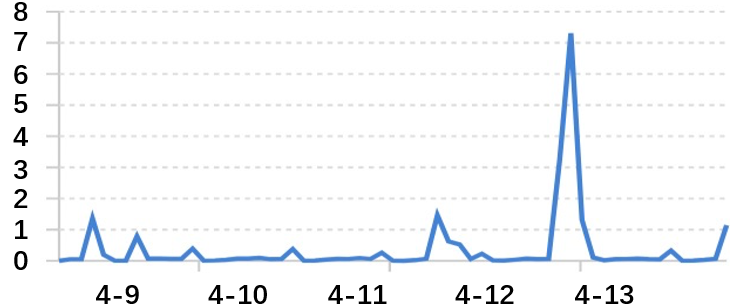}\label{fig:EC-week}
    }
    \subfigure[QPS fluctuations in one week]{
        \includegraphics[width=5.5cm]{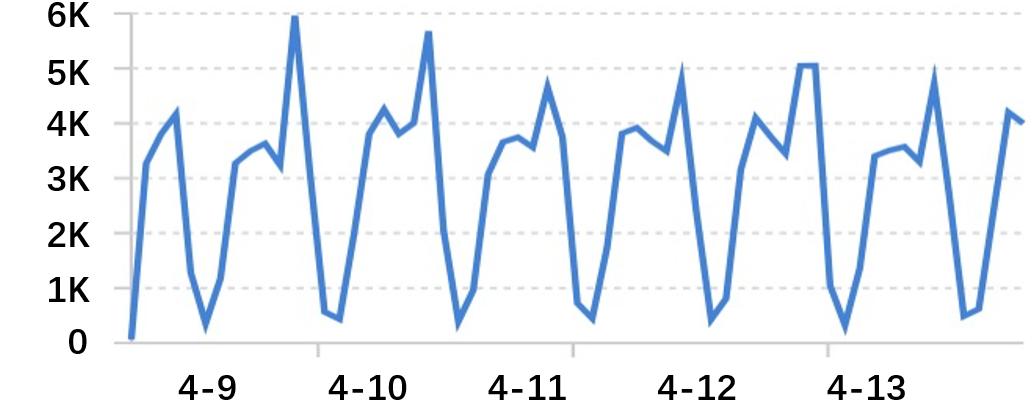}\label{fig:QPS-week}
    }
    \quad
    \caption{Fluctuation of Quality Metrics}\label{fig:all-metrics}
\end{figure*}

During the anomaly propagation chain analysis process, we need to continuously detect whether a service is an anomalous service in terms of certain quality metrics.
For example, for the analysis process shown in Figure~\ref{fig:propagation}, we first identify $S_4$ and $S_7$ in entry node analysis and then $S_7$, $S_9$, and $S_{10}$ in anomaly propagation chain extension by service anomaly detection.
Given an upstream/downstream service $S'$ of the current service $S$, we detect possible anomaly with $S'$ in terms of different anomaly types by analyzing the historical data of the corresponding quality metrics (i.e., RT, EC, QPS) of the service call between $S'$ and $S$.
Based on the characteristics of different quality metrics, we choose a different analysis model for each anomaly type.
These analysis models are designed based on the characteristics of the fluctuations of the corresponding quality metrics as shown in Figure~\ref{fig:all-metrics}.
These data are obtained from the monitoring system of Alibaba.


\subsubsection{Performance Anomaly}
\input{anomaly-RT}

\subsubsection{Reliability Anomaly}
\input{anomaly-EC}

\subsubsection{Traffic Anomaly}
\input{anomaly-QPS}

%% file: anomaly-RT.tex
Performance anomaly is detected based on the anomalous increase of response time (RT).
The challenge here is how to distinguish anomalous fluctuations from normal fluctuations. 
Figure~\ref{fig:RT-2-hour} and Figure~\ref{fig:RT-week}  illustrate the fluctuations of response time in two hours and in one week respectively.
It can be seen that there are periodic fluctuations in different periods of a day and different days of a week.
If we use intuitive rules like 3-sigma, it is likely that some of these periodic fluctuations will be recognized as performance anomalies.
To recognize anomalous fluctuations such as the anomalous point in Figure~\ref{fig:RT-2-hour}, we need to consider not only the quality metrics of the current period but also those of the same period of the previous day and the same day of the previous week.
Moreover, in historical data the response time is in normal fluctuations most of the time, and there are only a few anomalous fluctuations.

Based on the characteristics of RT, we choose to use OC-SVM (one class support vector machine) to train a prediction model for anomalous RT fluctuations.
OC-SVM~\cite{ocsvm} uses only the information for the target class (normal RT fluctuations) to learn a classifier that can recognize the samples belonging to the target class and identify the others as outliers (anomalous RT fluctuations).
OC-SVM is known to have the advantages of simple modelling, strong interpretability, and better generalization ability than clustering methods.
We define and use the following four kinds of features for the model.

\begin{itemize}
\item \textbf{Number of Over-Max Values}: the number of RT values in the current detection window that exceed the maximum RT value of a given comparison period.

\item \textbf{Delta of Maximum Values}: the delta between the maximum RT value in the current detection window and the maximum RT value of a given comparison period.

\item \textbf{Number of Over-Average Values}: the number of RT values in the current detection window that exceed the maximum moving average of the RT values of a given comparison period.

\item \textbf{Ratio of Average Values}: the ratio of the average of the RT values in the current detection window to the maximum moving average of the RT values of a given comparison period.
\end{itemize}

We consider the last 10 minutes as the current detection window, i.e., considering the 10 RT values of the last 10 minutes.
For the comparison periods, we consider the following three settings: the last one hour before the current detection window, the same hour of the previous day, the same hour of the same day of the previous week.
Therefore, we have 12 features in total for the OC-SVM model.

We use the Python based machine learning framework scikit-learn~\cite{scikitLearn} to implement the model.
To train the model, we collect 100,000 cases of different services from the monitoring infrastructure of Alibaba.
Each case includes a series of RT values of the current detection window and different comparison periods.
We also collect 600 cases for verifying the trained model. 
These cases are different from the training cases and the ratio of positive and negative cases is 1:1.
The results of the verification are shown in Table~\ref{tab:performance_train}, which provide the recall, precision,
F1-measure, FPR (False Positive Rate) of performance anomaly detection with different coverage of the training set.
It can be seen that the model can accurately detect performance anomalies with only 10-20\% of the training set, and it can achieve very high accuracy when using 70\% of the training set.

\begin{table}[ht]
	\newcommand{\tabincell}[2]{\begin{tabular}{@{}#1@{}}#2\end{tabular}}
	\caption{Accuracy of Performance Anomaly Detection}
	\label{tab:performance_train}
    \small
	\centering
		\begin{tabular}{|c|c|c|c|c|}
			\hline \textbf{Coverage} & \textbf{Recall} & \textbf{Precision} & \textbf{F1} & \textbf{FPR}\\
			\hline 10\%  &  0.86 & 0.73 & 0.79 & 0.32 \\
			\hline 20\%  &  0.81 & 0.88 & 0.84 & 0.12 \\
            \hline 30\%  &  0.82 & 0.92 & 0.87 & 0.07 \\
			\hline 40\%  &  0.83 & 0.92 & 0.87 & 0.08 \\
			\hline 50\%  &  0.84 & 0.94 & 0.89 & 0.06 \\
            \hline 60\%  &  0.83 & 0.94 & 0.88 & 0.06 \\
			\hline 70\%  &  0.86 & 0.95 & 0.90 & 0.05 \\
			\hline 80\%  &  0.87 & 0.95 & 0.91 & 0.05 \\
            \hline 90\%  &  0.87 & 0.95 & 0.91 & 0.04 \\
            \hline 100\% &  0.87 & 0.96 & 0.91 & 0.03 \\
			\hline
		\end{tabular}
	\vspace{-3pt}
\end{table}

%% file: anomaly-EC.tex
Reliability anomaly is detected based on the anomalous increase of error counts (EC).
As shown in Figure~\ref{fig:EC-2-hour} and Figure~\ref{fig:EC-week}, the value of EC is 0 most of the time.
In some cases, even when some errors are raised there is no influence on the business and the service may get back to normal in a short time.
For example, service calls fail and errors are raised when the circuit break is open, but the service will get back to normal soon after the load decreases and the circuit break is closed.
Therefore, we cannot use the performance anomaly detection model to detect reliability anomalies.
Binary classification models like Logic Regression (LR) or Random Forest (RF) can be used.
However, as only a small number of EC changes are reliability anomalies, LR models are likely to cause overfitting.

Based on the characteristics of EC, we choose to use Random Forest (RF) to train a prediction model for anomalous EC increases.
RF uses multiple decision trees for classification.
It can effectively combine multiple features and avoid overfitting.
We define and use the following five features for the model.
Note that some of the features combine other metrics (e.g., RT, QPS) together with EC, as anomalous EC increases often correlate with RT and QPS.
Similar to performance anomaly detection, we consider the last 10 minutes as the current detection window.


\begin{itemize}
\item \textbf{Previous Day Delta Outlier Value}: calculate the deltas of the EC values in the last one hour and the EC values in the same hour of the previous day; use the 3-sigma rule to identify possible outliers of the deltas in the current detection window; if exist return the average of the outliers as the feature value, otherwise return 0.

\item \textbf{Previous Minute Delta Outlier Value}: 
calculate the deltas of the EC values and the values of the previous minute in the last one hour; 
use the 3-sigma rule to identify possible outliers of the deltas in the current detection window; 
if exist return the average of the outliers as the feature value, otherwise return 0.

\item \textbf{Response Time Over Threshold}: whether the average RT in the current detection window exceeds a predefined threshold (e.g., 50ms).

\item \textbf{Maximum Error Rate}: the maximum error rate (i.e., EC divided by number of requests) in the current detection window.

\item \textbf{Correlation with Response Time}: the Pearson correlation coefficient (see Equation~\ref{eq:pearson}) between EC and RT in the current detection window.

\end{itemize}

We use the Python based machine learning framework scikit-learn~\cite{scikitLearn} to implement the model.
To train the model, we collect 1,000 labelled cases of different services from the monitoring infrastructure of Alibaba and the ratio of positive and negative samples is 1:3.
We also collect 400 cases for verifying the trained model and the ratio of positive and negative samples is 5:3.
The results of the verification are shown in Table~\ref{tab:reliability_train}, which provide the recall, precision,
F1-measure, FPR (False Positive Rate) of reliability anomaly detection with different coverage of the training set.
It can be seen that the model can accurately detect reliability anomalies with only 20\% of the training set, and it can achieve very high accuracy when using 40\% of the training set.

\begin{table}[ht]
	\newcommand{\tabincell}[2]{\begin{tabular}{@{}#1@{}}#2\end{tabular}}
	\caption{Accuracy of Reliability Anomaly Detection}
	\label{tab:reliability_train}
    \small
	\centering
		\begin{tabular}{|c|c|c|c|c|}
			\hline \textbf{Coverage} & \textbf{Recall} & \textbf{Precision} & \textbf{F1} & \textbf{FPR}\\
			\hline 10\%  &  0.29 & 1    & 0.44 & 0.70 \\
			\hline 20\%  &  0.64 & 1    & 0.78 & 0.35 \\
            \hline 30\%  &  0.64 & 1    & 0.78 & 0.35 \\
			\hline 40\%  &  0.86 & 0.97 & 0.91 & 0.14 \\
			\hline 50\%  &  0.88 & 0.95 & 0.91 & 0.12 \\
            \hline 60\%  &  0.88 & 0.90 & 0.89 & 0.12 \\
			\hline 70\%  &  0.93 & 0.90 & 0.92 & 0.09 \\
			\hline 80\%  &  0.95 & 0.95 & 0.94 & 0.07 \\
            \hline 90\%  &  0.95 & 0.95 & 0.95 & 0.04 \\
            \hline 100\% &  0.98 & 0.93 & 0.95 & 0.02 \\
			\hline
		\end{tabular}
	\vspace{-3pt}
\end{table}

%% file: anomaly-QPS.tex
Traffic anomaly is detected based on the anomalous fluctuations of queries per second (QPS).
As shown in Figure~\ref{fig:QPS-2-hour} and Figure~\ref{fig:QPS-week}, QPS complies with the normal distribution in both short term and long term.
Therefore, we choose to use the 3-sigma rule to detect anomalous QPS fluctuations.

Similar to performance and reliability anomaly detection, we consider the last 10 minutes as the current detection window.
Based on the QPS values in the last one hour, we use the 3-sigma rule to detect outliers in the current detection window.
To further eliminate false positives, we also check the Pearson correlation coefficient (see Equation~\ref{eq:pearson}) between the QPS values and the business metric values of the initial anomalous service.
Only when the correlation is higher than a predefined threshold (e.g., 0.9) and  3-sigma outliers are identified in the current detection window, a traffic anomaly is reported for the current service.


%% file: pruning.tex
An anomaly propagation chain may have multiple branches and the number of branches can continue to grow during anomaly propagation chain extension.
Therefore, a main challenge in anomaly propagation chain analysis lies in the exponential growth of the number of possible anomaly propagation branches.
In these branches, some anomalous services and service calls may be irrelevant to the reported availability issue.
To tackle the challenge and improve the efficiency of the analysis, we adopt a pruning strategy to eliminate irrelevant anomalous service calls.
The pruning strategy is based on the assumption that two successive edges (service calls) in an anomaly propagation chain have similar change trends of the corresponding quality metrics.
For example, in Figure~\ref{fig:propagation} the QPS of the edge $S_1 \rightarrow S_4$ should have similar change trend with the QPS of the edge $S_4 \rightarrow S_5$, otherwise the edge $S_1 \rightarrow S_4$ can be pruned.
Similarly, the RT of the edge $S_7 \rightarrow S_9$ and the RT of the edge $S_7 \rightarrow S_{10}$ should have similar change trends with the RT of the edge $S_5 \rightarrow S_7$.


Similar to the correlation estimation in candidate root cause ranking (see Section~\ref{sec:overview}), we use the Pearson correlation coefficient~\cite{perlationcorrelation} to measure the similarity of the change trends of the corresponding quality metrics of two successive service calls.
For each service call, the service call graph records a quality metric value (e.g., RT, EC, QPS) per minute.
To reflect the latest change trends, we only consider the quality metric values in a latest time window (e.g., the last 60 minutes).
The values for a specific quality metric of a service call in the last $n$ minutes thus can be represented by a vector $X$, where $X_i$ ($1 \leq i \leq n$) represents the value of the $i$th minute.
Given two service calls with the quality metric vectors $X$ and $Y$, their correlation coefficient can be calculated as Equation~\ref{eq:pearson}.

The pruning strategy is executed in the anomaly propagation chain extension process in the following way.
Before adding an anomalous node to the current anomaly propagation chain, check the correlation coefficient of the change trends of the corresponding quality metric of the current edge (service call) with the adjacent upstream/downstream edge.
If the correlation coefficient is lower than a threshold (e.g., 0.7), the current edge will be pruned and the current anomalous node will not be added to the current anomaly propagation chain.
For example, for the example shown in Figure~\ref{fig:propagation}, we will check the correlation coefficient of the QPS change trends of the edge $S_1 \rightarrow S_4$ with the edge $S_4 \rightarrow S_5$.
If the correlation coefficient is lower than the threshold, we will not add $S_1$ to the anomaly propagation chain even if it is a QPS anomalous node.
Similarly, we will check the correlation coefficients of the RT change trends of the edge $S_7 \rightarrow S_9$ and $S_7 \rightarrow S_{10}$ with the edge $S_5 \rightarrow S_7$.



%% file: experiment.tex
To evaluate the effectiveness and efficiency of \app~, we conduct a series of experimental studies to answer the following research questions.

\begin{itemize}
    \item \textbf{RQ1 (Localization Accuracy)}: How accurate is \app~for locating the root causes and anomaly types of availability issues of microservice systems?
    
    \item \textbf{RQ2 (Localization Efficiency)}: How efficient is the root cause localization process of \app? How well can it scale with the number of services?

    \item \textbf{RQ3 (Effect of Pruning)}: How does the pruning strategy influence the accuracy and efficiency of \app~with different similarity thresholds?
\end{itemize}

\subsection{Experimental Setup}
\input{setup}

\subsection{Localization Accuracy (RQ1)}
\input{rq1}

\subsection{Localization Efficiency (RQ2)}
\input{rq2}

\subsection{Effect of Pruning (RQ3)}
\input{rq3}

\subsection{Threats to Validity}
\input{threat}

%% file: setup.tex
We collect 75 availability issues from 28 subsystems of the e-commerce system of Alibaba.
These subsystems contain 265 services on average (min 7, max 1687, median 82).
The time of these availability issues ranges from February to June 2020.
All of them have been identified and annotated with anomaly types and root causes by operation engineers.
Each of them may have multiple anomaly types, and each anomaly type has a single root cause.
Among the 75 availability issues, 37 are caused by performance anomaly, 43 are caused by reliability anomaly, and 21 are caused by traffic anomaly.
The construction of the service call graph follows the practice in Alibaba, i.e., collecting service calls in the last 30 minutes and service call metrics (i.e., RT, EC, and QPS) in the last 60 minutes.


To evaluate the accuracy and efficiency, we compare \app~with the following two state-of-the-art baseline approaches.

\begin{itemize}
\item \textbf{MonitorRank}~\cite{cscs2013monitorRank}:
It detects root causes of anomalies in service-oriented systems and provides a ranked order list of possible root causes.
It uses the historical and current time-series metrics of each service along with service call graph to build an unsupervised model for ranking.


\item \textbf{Microscope}~\cite{ICSOC2018Microscope}:
It locates anomalous services with a ranked list of possible root causes in microservice systems.
It builds and uses service instance causality graph, which captures both communicating (service calls) and non-communicating (services co-located in the same machines) service dependencies.
It traverses the graph from the front-end service where an anomaly is reported and ranks possible root causes based the similarity of their metrics with those of the front-end service.
\end{itemize}


We use the top-$k$ hit ratio (HR@$k$) and mean reciprocal rank (MRR) to measure the accuracy of the root case localization.
\begin{itemize}

	\item HR@$k$ denotes the probability that the top $k$ result list contains the root cause.
	Let $A$ be the set of availability issues, $r_i$ be the root cause of the $i$th availability issue, $Rank_{i}^{k}$ be the top $k$ result list of the $i$th availability issue, HR@$k$ can be calculated as follows.

	\begin{equation}
	HR@k = \cfrac{1}{|A|}\sum\limits_{i=1}^{|A|}(r_i \in Rank_{i}^{k})
	\end{equation}

	\item MRR is the multiplicative inverse of the rank of the first correct answer.	
	If the correct answer is not included in the returned list, the rank can be regarded as positive infinity.
	Let $A$ be the set of availability issues and $Index_i$ be the rank of the root cause in the returned list of the $i$th availability issue, MRR can be calculated as follows.
	
	\begin{equation}
	MRR = \cfrac{1}{|A|}\sum\limits_{i=1}^{|A|}\cfrac{1}{Index_i}
	\end{equation}
	
\end{itemize}

All the experiments are conducted on a cluster with 15 virtual machines on Alibaba Cloud.
Each virtual machine is equipped with 4 Intel Xeon 2.50GHz CPU, 8 GB RAM and 60 GB Disk, and running Alibaba Group Enterprise Linux Server release 6.2.
The default settings of the thresholds are the following:
the RT threshold used in reliability anomaly detection is 50ms;
the correlation threshold used in the traffic anomaly detection is 0.9;
the correlation threshold used in the pruning strategy is 0.7.

%% file: rq1.tex
Table~\ref{tab:accuracy} shows the results of the overall accuracy evaluation of \app~and the two baseline approaches.
It can be seen that with \app~the top-1, top-3, and top-5 hit ratios are 0.48, 0.67, and 0.72, respectively, and the MRR is 0.58.
\app~significantly outperforms the baseline approaches in terms of all these metrics.

\begin{table}[ht]
	\newcommand{\tabincell}[2]{\begin{tabular}{@{}#1@{}}#2\end{tabular}}
	\caption{Overall Accuracy Evaluation (Best Scores are in Boldface)}
	\label{tab:accuracy}
    \small
	\centering
		\begin{tabular}{|c|c|c|c|c|c|}
			\hline \textbf{Metric} & \textbf{\app~} & \textbf{MonitorRank} & \textbf{Microscope}  \\
			\hline HR@1   & \textbf{0.48}  & 0.32   & 0.35  \\
			\hline HR@3   & \textbf{0.67}  & 0.40   & 0.49  \\
			\hline HR@5   & \textbf{0.72}  & 0.43   & 0.59    \\
			\hline MRR & \textbf{0.58}  & 0.37   & 0.44  \\
			\hline
		\end{tabular}
	\vspace{-3pt}
\end{table}

We also evaluate the accuracy of the three approaches for different anomaly types as shown in Table~\ref{tab:performance_algorithm_compare}.
It can be seen that \app~outperforms the two baseline approaches for all the three types of anomalies.
The advantages of \app~are the most significant for performance anomaly and the least significant for traffic anomaly.

\begin{table}[h]
	\newcommand{\tabincell}[2]{\begin{tabular}{@{}#1@{}}#2\end{tabular}}
	\caption{Accuracy Evaluation by Anomaly Types (Best Scores are in Boldface)}
	\label{tab:performance_algorithm_compare}
	\small
	\centering
		\begin{tabular}{|c|c|c|c|c|c|}
			\hline \textbf{Metric} & \textbf{\app~} & \textbf{MonitorRank} & \textbf{Microscope}\\
			\hline   \multicolumn{4}{|l|}{\cellcolor{gray!40} Performance Anomaly}  \\
			\hline HR@1    & \textbf{0.51} & 0.27  & 0.30   \\
			\hline HR@3    & \textbf{0.65} & 0.30  & 0.46   \\
			\hline HR@5    & \textbf{0.70} & 0.32  & 0.54   \\
			\hline MRR  & \textbf{0.61} & 0.31  & 0.39   \\
	   \hline   \multicolumn{4}{|l|}{\cellcolor{gray!40} Reliability Anomaly}  \\
			\hline HR@1    & \textbf{0.30 }& 0.26  & 0.26  \\
			\hline HR@3    & \textbf{0.56} & 0.33  & 0.44   \\
			\hline HR@5    & \textbf{0.70} & 0.35  & 0.52  \\
			\hline MRR  & \textbf{0.44} & 0.32  & 0.35  \\
			\hline   \multicolumn{4}{|l|}{\cellcolor{gray!40} Traffic Anomaly} \\
			\hline HR@1    & \textbf{0.46} & 0.32  & 0.36 \\
			\hline HR@3    & \textbf{0.55}  & 0.36  & 0.46  \\
			\hline HR@5    & \textbf{0.59} & 0.41  & 0.50  \\
			\hline MRR  & \textbf{0.54}  & 0.38  & 0.46 \\
			\hline
		\end{tabular}
	\vspace{-3pt}
\end{table}

The advantages of \app~can be explained by its specially designed mechanisms for service anomaly detection and anomaly propagation analysis.
MonitorRank considers the anomaly correlations between front-end services and back-end services and the correlations decrease with the propagation.
Therefore, it tends to choose services that are closer to the front-end service in the anomaly propagation chains.
In contrast, \app~considers the correlations between two successive service calls in the pruning.
Microscope uses the 3-sigma rule to detect anomalies and thus may produce false positives when occasional fluctuations occur.
In contrast, \app~uses more specific and sophisticated models for different types of anomalies


%% file: rq2.tex
\begin{figure}[ht]
	\centering
	\includegraphics[width=0.5\textwidth]{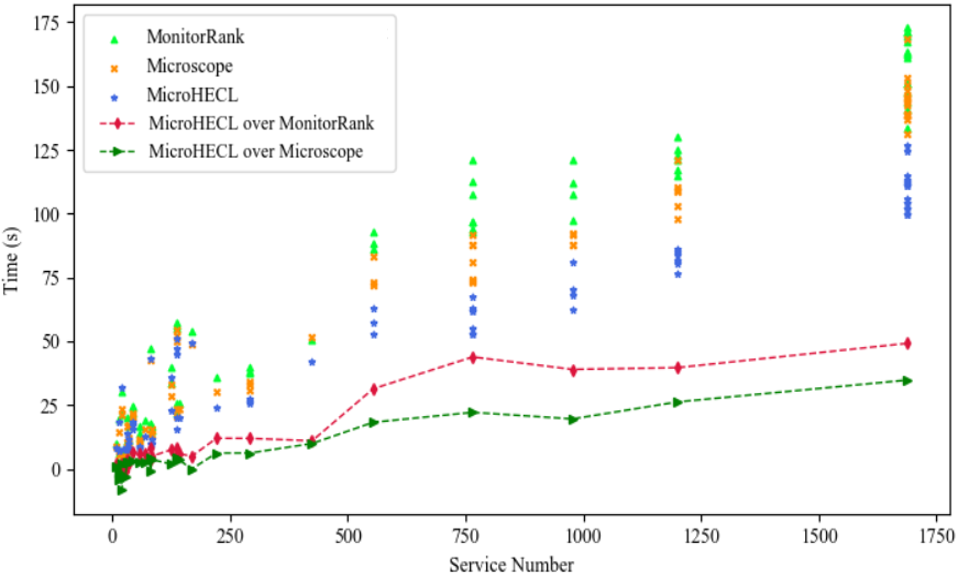}
	\vspace{-8pt}
	\caption{Detection Time Changes with Num of Nodes}\label{fig:time}
	\vspace{-5pt}
\end{figure}

The 75 availability issues are collected from 28 subsystems and these subsystems have different numbers of services.
To evaluate the efficiency and scalability of \app, we analyze the execution time of \app~and the two baseline approaches for the 75 availability issues and investigate how the time changes with the size (service number) of the target system.
The results are shown in Figure~\ref{fig:time}.
Note that there may be multiple availability issues collected from the same subsystem and each of them is indicated by a point.

In general, the execution time of \app~is 22.3\% less than that of Microscope and 31.7\% less than that of MonitorRank.
For the subsystems of different sizes \app~uses the least time among the three approaches for most availability issues.
The two curves in Figure~\ref{fig:time} show the changes of the time differences of \app~with the two baseline approaches respectively.
It can be seen that the advantage of \app~is not significant when the number of services is less than 250; when the number of services exceeds 250, the advantage of \app~is getting more and more significant with the increase of service number.
Moreover, the execution time of \app~(also the two baseline approaches) increases linearly with the increase of service number, showing a good scalability.

The advantages of \app~can be explained by its specially designed mechanisms for anomaly propagation analysis.
MonitorRank traverses the service call graph using random walk algorithm and needs to calculate the correlation scores for a large number of the nodes in the graph.
This process is time-consuming when the graph contains many nodes. 
Microscope traverses all the neighboring nodes of an anomalous node, both upstream and downstream.
Moreover, it has no pruning strategy.
In contrast, \app~considers only a single direction (upstream or downstream) for each anomaly type in anomaly propagation chain extension and uses a pruning strategy to eliminate branches that are likely irrelevant to the anomaly propagation.

%% file: rq3.tex
The pruning strategy is a key for the accuracy and efficiency of root cause localization.
We evaluate the effect of the pruning strategy by analyzing how the accuracy and time of root cause localization change with the threshold of correlation coefficient.
To this end, we run \app~to analyze the 75 availability issues with different threshold settings and measure the accuracy and time.
We choose the top-3 hit ratio (i.e., HR@3) as the indicator of accuracy.

\begin{figure}[ht]
	\centering
	\includegraphics[width=0.5\textwidth]{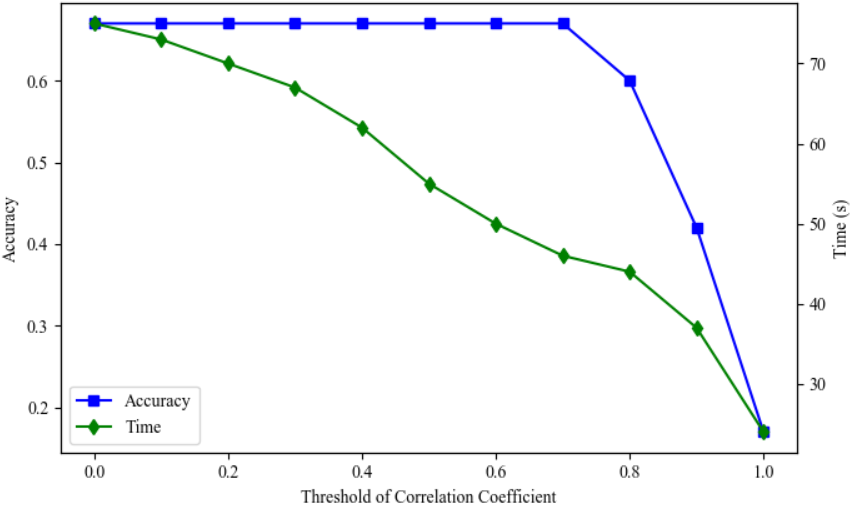}
	\vspace{-8pt}
	\caption{Evaluation of the Effect of the Pruning Strategy}\label{fig:rq3}
	\vspace{-10pt}
\end{figure}

The results of the evaluation are shown in Figure~\ref{fig:rq3}.
It can be seen that both the accuracy and time decrease with the increase of the threshold, as more services and service call edges are pruned in the analysis process and less services are reached and considered.
It can also be seen that the accuracy remains 0.67 when the threshold increases from 0 to 0.7, while the time decreases from 75 seconds to 46 seconds.
This analysis confirms the effectiveness of the pruning strategy, which can significantly improve the efficiency of root cause localization while keeping the accuracy.
And the best threshold is 0.7 for these availability issues.

%% file: threat.tex
A major threat to the internal validity of our studies lies in the implementation of the two baseline approaches.
We implement the two approaches by ourselves based on the descriptions in their papers and thus our implementations may not fully embody the approaches.
Another threat to the internal validity lies in the interactions with the monitoring infrastructure, i.e., EagleEye.
In the experiments, \app~together with the two baseline approaches need to obtain data from EagleEye in real time and there is thus uncertain latency due to limit circuit.
This threat applies to all the three approaches equally.

A major threat to the external validity of our studies lies in the fact that all the studies are conducted with the data and availability issues collected from the e-commerce system of Alibaba.
Microservice systems of other companies or domains may have different characteristics, for example, in the fluctuations of quality metrics and propagation of service anomalies.
The accuracy and efficiency achieved in our studies may not apply to other microservice systems.

%% file: application.tex
\app~has been implemented and deployed as a distributed system in Alibaba based on EagleEye and other infrastructures.
To support finer-grained localization, the system also considers local middleware components (including databases, message queues, caches) of services as the targets of localization.
These components and their interactions with services are incorporated into the service call graph and the quality metrics of the interactions are also recorded for analysis.

\app~has been deployed in Alibaba for more than 5 months and used to handle more than 600 availability issues.
For each availability issue the system recommends the top-3 candidate root causes for the developers.
Before the application of \app, the developers needs to use the service call graph visualization and quality metrics analysis provided by the web interfaces of EagleEye to manually identify possible root causes.
For each availability issue, the typical time of root cause localization and confirmation is over 30 minutes.
With \app, the system can produce root cause recommendations in 76 seconds on average and the developers usually can confirm the results and start fault repairing and recovery in 5 minutes after an availability issue is detected.
The hit ratio (i.e., HR@3) of \app~is 68\%.
Feedback from the developers shows that most of them trust the recommendations of the system and treat the recommendations as the root causes by default.

%

We analyze the cases for which \app~cannot accurately recommend the root causes and find that the approach and the system need to be improved from multiple aspects.
First, it is possible that availability issues are detected but there are no anomalies with the quality metrics of services.
Second, anomalous fluctuations may accumulate slowly and exceed the predefined time window.
For example, an anomalous fluctuation may start more than one hour before an availability issue is detected, thus the anomalous fluctuation may not be detected by considering the changes of quality metrics in the last one hour.
More advanced techniques are required to improve the current approach.



%% file: related.tex

Traditional anomaly detection approaches for distributed and cloud-based systems are mainly based on log analysis.
Fu \emph{et. al}\cite{09ICDM_EADULA}  proposed an unstructured log analysis technique of anomalies detection for distributed systems. 
Xu \emph{et al.}\cite{17ICEBE_LogDC} introduced LogDC, a log model based problem diagnosis tool for cloud applications with the full-lifecycle Kubernetes logs.
Jia \emph{et al.}\cite{17ICWS_ADHGML} proposed an approach for automatic anomaly detection based on logs. It raises anomaly alerts on observing deviations from the hybrid model which captures normal execution flows inter and intra services.
These approaches build anomaly detection modes with log parsing.
They cannot detect the anomalies with quality metrics of services, nor can they support anomaly propagation analysis for root cause localization.

Root cause analysis in practice often relies on visualization of logs and traces.
Zhou \emph{et al.}~\cite{Zhou_TSE18MS} presented an improved visualization analysis approach for fault analysis and debugging of microservice systems.
Wang \emph{et al.}~\cite{Hanzhang_GraphBased2019} demonstrated Grano, an end-to-end anomaly detection and root cause analysis system for cloud native distributed data platform by providing a holistic view of the system component topology, alarms and application events.
Guo \emph{et al.}~\cite{Guo_FSE20Industry} proposed a graph-based trace analysis approach for understanding architecture and diagnosing problems of microservice systems.
There approaches the developers to manually detect the root causes with the aid of trace and log visualization.

Recently, trace analysis based approaches have been proposed for automatic root cause localization for microservice or service-based systems.
Pham \emph{et al.}\cite{Cuong_traceBased} introduced a fault localization approach based on trace similarity comparison. It collects a large number of traces labelled with fault types and root causes through fault injection and conducts fault localization by comparing between faulty traces and historical traces.
Gan \emph{et al.}\cite{19ICASPLOS_Seer} presented Seer, an online cloud performance debugging system for QoS violations by learning spatial and temporal patterns from traces.
Zhou \emph{et al.}\cite{FSE2019LatentError} proposed MEPFL, an approach that can predict latent errors and locate root causes for microservice applications by learning from system trace logs.
It predicts latent errors, faulty microservices, and fault types for trace instances captured in the production environment using models trained based on a set of features defined on microservice traces.
These approaches often need to collect and analyze a large number of traces to extract anomaly patterns or train prediction models, thus are not efficient for large-scale microservice systems.

Some researchers have proposed approaches that use service call graphs or causality graphs for automatic root cause localization for microservice or service-based systems.
{\'{A}}lvaro \emph{et al.}~\cite{GraphBasedformicroservice} presented a root cause analysis framework, based on graph representations of service-oriented and microservice architectures. It analyzes anomalies by comparing the similarity between the fault subgraph and the historical subgraph.
Li \emph{et al.}~\cite{NOMS2020MicroRCA} proposed MicroRCA, a system to locate root causes of performance issues in microservice systems. MicroRCA infers the root causes by correlating the performance symptoms with the corresponding resource utilization based on an attributed graph.
Wang \emph{et al.}~\cite{DCCGRID2018CloudRange} proposed CloudRanger, a system dedicated for cloud native systems. CloudRanger identifies the culprit services which are responsible for cloud incidents by constructing the impact graph among services with a dynamic causal relationship analysis approach. 
Kim \emph{et al.}~\cite{cscs2013monitorRank} introduced MonitorRank, an algorithm that can reduce the time, domain knowledge, and human effort required to find the root causes of anomalies in service-oriented architectures. MonitorRank finds root causes with service call graph and historical and current time-series metrics of services.
Lin \emph{et al.}~\cite{ICSOC2018Microscope} proposed Microscope, a system to identify and locate the anomalous services with a ranked list of possible root causes in microservice environments. Microscope infers the causes of performance problems in real time by constructing a service causal graph.
The preceding approaches share some similar ideas with ours.
However, most of these approaches detect anomalous services by simply analyzing the correlations between metrics or statistical properties and do not consider the characteristics of different anomaly types and quality metrics. 
Moreover, these approaches need to traverse most of the nodes in the graph, thus may be inefficient for large-scale systems.
In contrast, we combine machine learning and statistical methods and design a customized model for the detection of each type of anomalies and adopt a pruning strategy to improve the efficiency of anomaly propagation chain analysis.

%% file: conclusion.tex
In this paper, we propose a high-efficient root cause localization approach for availability issues of microservice systems, called \app.
It dynamically constructs a service call graph and analyzes possible anomaly propagation chains by traversing the graph along anomalous service calls.
Different from existing approaches, \app~uses customized models based on machine learning and statistical methods to detect different types of service anomalies (i.e., performance, reliability, traffic) and employs a pruning strategy to eliminate irrelevant service calls in anomaly propagation chain analysis.
The accuracy and efficiency of \app~have been confirmed by both experimental studies and the practical application in Alibaba.
Future work will be focused on efficiently combing traces, logs, and metrics for more accurate and explainable root cause localization.